\documentclass[aps,prd,amsmath,amssymb,reprint,nofootinbib]{revtex4-1}

\usepackage[utf8]{inputenc}
\usepackage{amsmath}
\usepackage{color}
\usepackage[breaklinks]{hyperref}
\usepackage{color}
\usepackage{subfigure}
\usepackage{verbatim}
\usepackage{subfigure}
\usepackage{graphicx}
\usepackage[normalem]{ulem}

\begin{document}
\title{Stability of the  de-Sitter spacetime. \\
 The anisotropic case.
}
\author{Przemysław Bieniek, Jan Chojnacki, Jan H. Kwapisz, Krzysztof A. Meissner}
\affiliation{Faculty of Physics, University of Warsaw,\\
Pasteura 5, Warsaw, Poland}

\begin{abstract}
We elaborate on the cosmological implications of the recently established non-perturbative $O(D,D)$-symmetric approach. Following the $\alpha'$-complete Friedmann equations, we provide a qualitative description of the dynamics of anisotropy perturbations. We find that the spacetime stability depends on the sign of the generalized dilaton time derivative. We consider isotropic limits and find possible mechanism for pre-big bang cosmology, goverend by the value of cosmological constant.

\end{abstract}


\maketitle




\section{Introduction}

The discovery of the positive cosmological constant has put the existence of de-Sitter solutions and their stability as one of the central research questions within the string theory community. The so-called de-Sitter conjecture states that the de-Sitter vacua are unstable within string theory \cite{Obied:2018sgi, Garg:2018reu, Ooguri:2018wrx}. 

On the other hand a suitable $O(D,D)$ formalism is at hand to study the de-Sitter solutions. As found by K.A. Meissner and G. Veneziano \cite{Meissner:1991zj, MEISSNER_1991}, in $D+1$ dimensions the low-energy stringy effective action once reduced to time-dependent backgrounds possess continuous $O(D,D)$ symmetry, see also \cite{Sen:1991zi, Maharana:1992my}. The symmetry emerges in the massless sector of (super) string theories: scalar dilaton $\phi(x)$, the spin-2 graviton $G_{\mu\nu}$ and anticommuting Kalb-Ramond 2-form field $B_{\mu\nu}$. The low-energy action takes the form
\begin{align}
\label{EffectiveAction}
    I&=\frac{1}{2\kappa^2}\int d^{1+D} x\, \sqrt{-G} e^{-2\phi}(R+4G^{\mu\nu}\partial_\mu \phi \partial_\nu \phi\nonumber\\
    &-\frac{1}{12}H_{\mu\nu\rho} H^{\mu\nu\rho} ),
\end{align}
where $D$ is the number of spatial dimensions, $R$ is the Ricci scalar curvature of the metric $G_{\mu\nu}$ and $H_{\mu\nu\rho}$ is a field strength of $B_{\mu\nu}$, defined as $H_{\mu\nu\rho}:=\partial_\mu B_{\nu\rho}+\partial_\nu B_{\rho\mu}+\partial_{\rho} B_{\mu\nu}$.
For the time-dependent backgrounds $G_{\mu\nu} =: g_{\mu\nu}(t)$, $B_{\mu\nu}=:b_{\mu\nu}(t)$, $\phi=:\phi(t)$ the effective action (\ref{EffectiveAction}) may be expressed in the manifestly $O(D,D)$ invariant form:
\begin{align}
\label{invariant action}
    I \varpropto \int dt\, e^{-\Phi} \left[-\dot{\Phi}^2-\frac{1}{8} \textrm{Tr}\left(\mathcal{\dot{S}}^2\right) \right],
\end{align}
We call $\Phi=2\phi-\ln\sqrt{\det g} $ the generalized (shifted) dilaton field and $g$ is the spatial metric tensor. Following the formalism of \cite{Meissner:1991zj}, $2d$-dimensional matrix $\mathcal{S}$ is defined:
\begin{equation}\label{eq:S definition}
\mathcal{S} = \begin{pmatrix}
			bg^{-1} & g - bg^{-1}b\\
				g^{-1} & -g^{-1}b
			\end{pmatrix}.\\  
\end{equation}
The global $O(D,D)$ group acts as
\begin{align}
\label{eq:Odd transformation}
    \Phi \xrightarrow{}\Phi,\quad \mathcal{S}\xrightarrow{}\Omega ^T \mathcal{S}\Omega.
\end{align}
As shown by \cite{Meissner1997} also for the $\alpha'$ corrected actions the fields can be transformed in such a way that $O(D,D)$ symmetry is manifest. Using this reasoning, in  \cite{Hohm:2019jgu}, the full, non-perturbative form of the $\alpha'$ corrected effective action on cosmological backgrounds has been derived 
\begin{align}\label{eq:hohmzwiebachaction}
		I = \int\,dt\,e^{-\Phi} \left[-\dot{\Phi}^2 + \sum_{k=1}^{\infty} \alpha'^{k-1}\left(c_k \mathrm{Tr}\left(\dot{\mathcal{S}}^{2k}\right) \right.\right.\\
		+ \textrm{multitrace}\Bigl) \Biggr].
	\end{align}
A generic multitrace term at order $\alpha'^{k-1}$ is of the following form:
\begin{equation}\label{eq:multitrace}
    \prod_i\mathrm{Tr}\left(\dot{\mathcal{S}}^{2n_i}\right)
\end{equation} 
with $\sum_in_i = k$ and $n_i > 1$. Note that also a cosmological constant term may be added in the $O(D,D)$ manner, \cite{2020_Brandenberger}. The higher order potential terms in $\Phi$ will be $O(D,D)$ symmetric as well, but will cause the instability in the equations \cite{Bieniek:2022mrv}.

This general form of the string theory effective action and the associated equations revealed a new possibility of a non-perturbative de-Sitter solution once summing all the possible $\alpha'$ corrections in the $O(D, D)$ formalism \cite{Hohm:2019jgu}, see also other developments here \cite{2020_Brandenberger, Wang:2019kez, Hohm:2019ccp, 2020_Bernardo, 2021Nunez, Basile:2021euh, Basile:2021krk, Basile:2021krr}. Finally, in our previous article, we studied the isotropic case for constant shifted dilaton $\dot{\Phi}=0$ 
\cite{Bieniek:2022mrv} as well as in the presence of cosmological constant.

In this work, we extend our study towards the anisotropic case and the stability of the de-Sitter solution across different spatial dimensions. As it will turn out, the $O(D, D)$ framework will not be suitable to study the anisotropies, thus here we have adopted the perturbative approach including the multitrace effects. We also comment on possible Pre-Bing Bang scenarios.

\section{Anisotropic perturbations}
\label{sec:anisotropy}
The anisotropic solutions for the full $\alpha'$ $O(d, d)$ dynamics have been discussed in 
\cite{Bernardo:2020nol, 2021Nunez}. Yet in both of the articles only a special class of the anisotropies have been covered, namely for the $d-n$ static dimensions, \textit{i.e.} $H_i=0$ for $i \in (n, d)$. As we will shortly discuss, the full unperturbative, dynamical investigation of anisotropic perturbations within Hohm-Zwiebach setup is beyond the feasible scope of the theory. We are however able to recognise stability features present in the isotropic non-perturbative case.\\
To investigate the stability of the $d$ dimensional de-Sitter solution, let us now consider the anisotropic case:
\begin{align}
\label{eq:Anisotropic}
	g_{ij} & = \mathrm{diag}\left(a_1^2(t), a_2^2(t), \ldots, a_d^2(t)\right) = \mathbf{a}^2, &
	b_{ij} &= 0.
\end{align}
In that case, the matrix $\mathcal{S}$ equals:
\begin{equation}
	\mathcal{S} = \begin{pmatrix}
			  		0 & g \\
			  		g^{-1} & 0
			  	  \end{pmatrix}
		  	  	= \begin{pmatrix}
		  	  		0 & \mathbf{a}^{2}\\
		  	  		\mathbf{a}^{-2} & 0
		  	  	\end{pmatrix}.
\end{equation}
And its derivative is:
\begin{equation}
	\dot{\mathcal{S}} = \begin{pmatrix}
							0 & 2\mathbf{a}\dot{\mathbf{a}}\\
							-2\mathbf{a}^{-3}\dot{\mathbf{a}} & 0
						\end{pmatrix}.
\end{equation}
Then it can be seen that the square of the derivative takes a diagonal form:
\begin{equation}\label{eq:S squared}
	\dot{\mathcal{S}}^2 = -4\begin{pmatrix}
		 						\mathbf{H}^2& 0\\
								0 & \mathbf{H}^2
							\end{pmatrix},
\end{equation}
where we define $\mathbf{H} = \dot{\mathbf{a}}\mathbf{a}^{-1} = \mathrm{diag}(H_1, H_2, \ldots, H_d)$.

In \cite{Hohm:2019jgu}, a simpler case of $a_1 = a_2 = \ldots = a_d$ is considered. It is then shown that in the low-energy effective action the multitrace terms can be ignored, because they contribute exclusively to the value of the constants $c_k$. It is because then (as one can see by setting all $H_i$ to be equal in \eqref{eq:S squared}):
\begin{equation}
	\mathrm{Tr}(\dot{\mathcal{S}}^{2l}) = (-1)^l2^{2l+1}dH^{2l}.
\end{equation}
Since the multitrace terms at order $\alpha'^{k-1}$ consist of the traces of powers of $\dot{\mathcal{S}}$ that sum to $2k$, they will all have the form $CH^{2k}$, where $C$ is constant. Therefore, they will all sum to a term proportional to $H^{2k}$ and can be included by simply changing the constants $c_k$.

However, the anisotropic case is different. A generic trace appearing in the action is:
\begin{align}\label{eq:anisotropic trace}
	\mathrm{Tr}(\dot{\mathcal{S}}^{2l}) = (-4)^{l}\mathrm{Tr}\begin{pmatrix}
											\mathbf{H}^{2l}& 0\\
											0 & \mathbf{H}^{2l}
											\end{pmatrix}
	=(-4)^{l}2\sum_{i=1}^{\infty}H_i^{2l}.
\end{align}
Plugging this into \eqref{eq:multitrace}, it is easily seen that the multitraces will introduce terms proportional to 
\begin{align}
\prod_{i}H^{2n_i}_{j_i}
\end{align}
 where some of the $j_i$ are not equal, into the action. The consequence of this is that the equations become hard to derive, and they would be a lot more complicated than in the isotropic case.

However, in \cite{Hohm:2019jgu}, it has been shown that the first multitrace terms will appear at the third order in $\alpha'$. Therefore we will disregard the multitrace terms if we truncate the expressions to a sufficiently small order. Alternatively, the issue of existence of de-Sitter vacua can be investigated by assuming, that the matrices $g$, $\dot{g}$, and $\dot{b}$ commute (the so called commuting matrices ansatz \cite{Nunez:2020hxx}). That approach allows for some non-perturbative analysis, but results in very complicated equations which are not suitable for stability analysis. In our study, we take one another perspective: the perturbative approach quantifying multitrace effects affecting the stability of a general anisotropic solution.\\
The equations derived in \cite{Hohm:2019jgu} by a direct variation of the action (without the multitrace terms) with the added contribution from the cosmological constant are:
\begin{align}
 \sum_{k=1}^{\infty}\alpha'^{k-1}kc_k\left(\frac{d}{dt}\dot{\mathcal{S}}^{2k-1}-\dot{\Phi}\dot{\mathcal{S}}^{2k-1}+\mathcal{S}\dot{\mathcal{S}}^{2k}\right) & = 0, \label{eq:ES}\\
	  \ddot{\Phi}-\sum_{k=1}^{\infty}\alpha'^{k-1}kc_k\mathrm{Tr}(\dot{\mathcal{S}}^{2k}) & = 0, \label{eq:EPhi}\\
	  \dot{\Phi}^2-\sum_{k=1}^{\infty}\alpha'^{k-1}(2k-1)c_k\mathrm{Tr}(\dot{\mathcal{S}}^{2k}) - \Lambda & = 0 .\label{eq:En}
\end{align}
They follow from variation with respect to $\mathcal{S}$, $\Phi$, and the lapse function $n(t)$, which has been set to 1.
In the considered case, the equation \eqref{eq:En} transforms into:
\begin{align}
		\dot{\Phi}^2 + \sum_{k=1}^{\infty}\left((-\alpha')^{k-1}2^{2k+1}(2k-1)c_k\sum_{i=1}^{d}H_i^{2k}\right) - \Lambda& = 0,		
	\end{align}
the following relation between $\dot{\Phi}^2$ and $\ddot{\Phi}$ holds:
\begin{align}
		\dot{\Phi}^2 + \sum_{k=1}^{\infty}\left((-\alpha')^{k-1}2^{k+1}(k-1)c_k\sum_{i=1}^{d}H_i^{2k}\right) - \Lambda& = \ddot{\Phi},		
	\end{align}
and \eqref{eq:ES} gives $d$ equations of the form:
\begin{equation}\label{eq:EHi}
	\left[\frac{d}{dt}-\dot{\Phi}\right]\sum_{k=1}^{\infty}(-\alpha')^{k-1}2^{2k-2}kc_kH_i^{2k-1} = 0.
\end{equation}
\subsection{The perturbative approach}
Now, we proceed to the calculation of the terms introduced to the equations by the first multitrace term. The part of the action that includes it is:
\begin{equation}
    I^{(m)} = \int\,dt e^{-\Phi} \alpha'^3c_{4,1}\left(\mathrm{Tr}\left(\dot{\mathcal{S}}^4\right)\right)^2.
\end{equation}
In order to vary it with respect to $\mathcal{S}$, we use an approach laid out in \cite{Hohm:2019jgu}. First, we perform a variation as if $\mathcal{S}$ was an unconstrained variable:
\begin{align}
    \delta_{\mathcal{S}}I_m &= \int\,dte^{-\Phi}8\alpha'^3c_{4,1}\mathrm{Tr}\left(\dot{\mathcal{S}}^4\right)\mathrm{Tr}\left(\frac{d\delta\mathcal{S}}{dt}\dot{\mathcal{S}}^3\right)\notag\\
    &= 8\alpha'^3c_{4,1}\int\,dte^{-\Phi}\mathrm{Tr}\left(\dot{\mathcal{S}}^4\right)\mathrm{Tr}\left(\frac{d}{dt}\left(\delta\mathcal{S}\dot{\mathcal{S}}^3\right)-\delta\mathcal{S}\frac{d}{dt}\dot{\mathcal{S}}^3\right) \notag\\
    &= 8\alpha'^3c_{4,1}\int\,dte^{-\Phi}\mathrm{Tr}\left(\delta\mathcal{S}\left(\mathrm{Tr}\left(\dot{\mathcal{S}}^4\right)\dot{\Phi}\dot{\mathcal{S}}^3\right.\right. \nonumber \\
  &  \left.\left.
  -\frac{d}{dt}\mathrm{Tr}\left(\dot{\mathcal{S}}^4\right)\dot{\mathcal{S}}^3 -\mathrm{Tr}\left(\dot{\mathcal{S}}^4\right)\frac{d}{dt}\dot{\mathcal{S}}^3\right)\right).
\end{align}
The expression multiplying $\delta\mathcal{S}$ defines $F_{\mathcal{S}}^{(m)}$. The contribution to equation \eqref{eq:ES} is given by 
\begin{align}
E_{\mathcal{S}}^{(m)} = \frac{1}{2}(F_{\mathcal{S}m}-\mathcal{S}F_{\mathcal{S}m}\mathcal{S}). 
\end{align}
Using $\mathcal{S}^2=1$, $\mathcal{S}\dot{\mathcal{S}}=-\dot{\mathcal{S}}\mathcal{S}$ and $\mathcal{S}\frac{d}{dt}\dot{\mathcal{S}}^{2k-1}\mathcal{S} = -\frac{d}{dt}\dot{\mathcal{S}}^{2k-1}-2\mathcal{S}\dot{\mathcal{S}}^{2k}$, it can be brought to the following form:
\begin{align}
     E_{\mathcal{S}}^{(m)} =-8\alpha'^3c_{4,1}\left[\mathrm{Tr}\left(\dot{\mathcal{S}}^4\right)\frac{d}{dt}\dot{\mathcal{S}}^3+\mathrm{Tr}\left(\dot{\mathcal{S}}^4\right)\mathcal{S}\dot{\mathcal{S}}^4\right.\nonumber \\
 \left.   -\left(\mathrm{Tr}\left(\dot{\mathcal{S}}^4\right)\dot{\Phi}-\frac{d}{dt}\mathrm{Tr}\left(\dot{\mathcal{S}}^4\right)\right)\dot{\mathcal{S}}^3\right].
\end{align}
Using \eqref{eq:anisotropic trace}, this can be shown to give the following corrections to equations \eqref{eq:EHi}:
\begin{align}
    E_{H_i}^{(m)}=2^{11}\alpha'^3c_{4,1}\left[3\left(\sum_j H_j^4\right)\dot{H_i}H_i^2\right.\nonumber\\
    \left.-\left(\sum_j H_j^4\right)\dot{\Phi}H_i^3+4\left(\sum_j \dot{H_j}H_j^3\right)H_i^3\right].
\end{align}
Additionally, the equations \eqref{eq:EPhi} and \eqref{eq:En} will receive contributions from the multitrace term. For our purposes, only the change to the latter needs to be calculated. After restoring $g_{00} = -n(t)$, the multitrace term will be multiplied by $\frac{1}{n^7}$. Therefore, its contribution to the equation will be:
\begin{align}
    E_{n}^{(m)} = -7\alpha'^3c_{4,1}\left(\mathrm{Tr}\left(\dot{\mathcal{S}}^4\right)\right)^2 = -7\cdot2^{10}\alpha'^3c_{4,1}\left(\sum_j H_j^4\right)^2.
\end{align}
Finally, the equations to the third order take the form:

\begin{align}
	-\left[\frac{d}{dt}-\dot{\Phi}\right](4c_1H_i -32\alpha'c_2H_i^3 \nonumber\\
	+192\alpha'^2c_3H_i^5-2^{10}\alpha'^3c_{4,0}H_i^7)+ E_{H_i}^{(m)}&=0, \label{eq:Hi equations}\\
	\dot{\Phi}^2 + 8c_1\sum_{j=1}^{d}H_j^{2} - 96\alpha'c_2\sum_{i=j}^{d}H_j^{4} \nonumber\\
	+5\cdot2^7\alpha'^2c_3\sum_{i=j}^{d}H_j^6-7\cdot2^9\alpha'^3c_{4,0}\sum_{i=j}^{d}H_j^8+\Lambda+E_{n}^{(m)}&=0. \label{eq:dotPhi squared equation}
\end{align}
Using equation \eqref{eq:dotPhi squared equation}, $\dot{\Phi}$ can be expressed in terms of $H_i$:
\begin{widetext}
\begin{equation}\label{eq:dotPhi equation}
	\dot{\Phi} = \mathrm{sgn}(\dot{\Phi})\sqrt{-8c_1\sum_{j=1}^{d}H_j^{2} + 96c_2\alpha'\sum_{j=1}^{d}H_j^{4}-5\cdot2^7c_3\alpha'^2\sum_{j=1}^{d}H_j^6+7\cdot2^9c_{4,0}\alpha'^3\sum_{i=j}^{d}H_j^8-E_{n}^{(m)} + \Lambda}.
\end{equation}
\end{widetext}

By plugging \eqref{eq:dotPhi equation} into \eqref{eq:Hi equations}, we get $d$ differential equations for the $H_i$'s, where the only appearance of $\Phi$ is the sign of its derivative. They are linear equations for $\dot{H_i}'s$ and can be used to express them in terms of $H_i$ and $\mathrm{sgn}(\dot{\Phi})$.
Time evolution of $\dot\Phi$ is entirely determined by the evolution of anisotropic Hubble parameter. We note that the sign of the generalised dilaton time derivative, is governed by the equation of motion containing the second time derivative, following from \eqref{eq:EPhi}. Moreover, we should consider such constants $c_n$, so that the expression under the squere root is positively defined or find a dynamical mechanism restricting the flow into this area. We discuss such mechanism in section \ref{sec:perturbative}.
\section{Exact solution to zero-order anisotropic equations}\label{app:exact solution}

If we disregard the $\alpha'$ terms in the system of equations \eqref{eq:ES}, \eqref{eq:EPhi} and \eqref{eq:En} for the case considered in section \ref{sec:anisotropy} with $\Lambda = 0$, they can be solved analytically. We consider the cases $\Lambda = 0$ and $\Lambda \neq 0$ separately. 

\subsection{Zero potential}
In this case, \eqref{eq:EPhi} and \eqref{eq:En} imply the following relation:
\begin{equation}
	\dot{\Phi}^2=\ddot{\Phi}.
\end{equation}
The general solution for $\Phi$ is given by:
\begin{align}
	\Phi(t)& = -C_1\mathrm{log}(t-t_0)+\Phi_0\label{eq:Phi interesting solution},
\end{align}
with $C_1 \in \{0,1\}$. For $C_1=0$ the Eq.\eqref{eq:ES} implies that $H_i$ are also constant. For $C_1 = 1$, after plugging \eqref{eq:Phi interesting solution} into \eqref{eq:ES}, one gets:
\begin{equation}
	\dot{H_i}+\frac{1}{t-t_0}H_i=0.
\end{equation}
They can be solved to give:
\begin{equation}\label{eq:Hi solution}
	H_i(t) = \frac{\beta_i}{t-t_0},
\end{equation}
where $\beta_i$ are constants. By plugging \eqref{eq:Hi solution} and \eqref{eq:Phi interesting solution} into \eqref{eq:En}, a constraint on $\beta_i$'s is:
\begin{equation}
    \sum_j \beta_j^2 = 1.
\end{equation}
Now, since $H_i = \frac{\dot{a}_i}{a_i}$, we solve for $a_i$ and obtain:
\begin{equation}
	a_i(t) = D_i(t-t_0)^{\beta_i}.
\end{equation}
And since $\phi = \frac{1}{2}(\Phi+\mathrm{log}(\sqrt{\mathrm{det}g}))$ and $\sqrt{\mathrm{det}g} = \prod_j a_j$, we have:
\begin{equation}
	\phi(t) = \frac{1}{2}\left(\sum_j\beta_j-1\right)\mathrm{log}(t-t_0)+\phi_0.
\end{equation}
 Interestingly, for $\sum_j\beta_j=1$ the dilaton remains constant for time dependent $g$. The dilaton and Ricci scalar curvature are singular for $t\xrightarrow{}t_0$. This means that the action is infinite at this limit and in the spirit of Finite Action Principle \cite{BarrowTipler,Barrow:2019gzc,Lehners_2019,borissova2020blackhole,Jonas_2021,chojnacki2021finite} such solution will be dynamically excluded, as the exponential weight in the path integral will vanish.
There is however, an exception to this. Notice, when $\sum \beta_i=1$ the dilaton and curvature remain finite. Assuming all $|\beta_i|$ are equal, the only possible choice of $\beta_i$ is $\{-\frac{1}{3},-\frac{1}{3},-\frac{1}{3},+\frac{1}{3},\dots,\frac{1}{3},\}$ corresponding to six contracting dimensions and three expanding. Originally, this result was discovered by Meissner and Veneziano in \cite{MEISSNER_1991}. At higher orders of $\alpha'$, the exact solution is much harder to obtain, since terms without $\Phi$ in \eqref{eq:EPhi} and \eqref{eq:En} differ and \eqref{eq:Phi interesting solution} doesn't hold.
\subsection{Cosmological constant potential}

Now, \eqref{eq:EPhi} together with \eqref{eq:En} give us:
\begin{equation}
\label{first-order-dilaton-evolution}
    \dot{\Phi}^2=\ddot{\Phi} + \Lambda.
\end{equation}
This equation has two solutions:
\begin{align}
    \Phi_1(t) &= -\mathrm{log}\left[\mathrm{cosh}\left(\sqrt{\Lambda}\left(t_0-t\right)\right)\right] + \Phi_0;\\
    \Phi_2(t) &= -\mathrm{log}\left[\mathrm{sinh}\left(\sqrt{\Lambda}\left(t_0-t\right)\right)\right] + \Phi_0;\\
    \Phi_3(t) &= \pm\sqrt{\Lambda}t + \Phi_0.
\end{align}
Then, similarly as before, we can obtain solutions for $a_i$:

\begin{align}
    a_{1i}(t) &= \mathrm{exp}\left[2\gamma_i\mathrm{arctan}\left(\mathrm{tanh}\left(\frac{\sqrt{\Lambda}\left(t_0-t\right)}{2}\right)\right)\right];\label{eq:cc_solution1}\\
    a_{2i}(t) &= \mathrm{tanh}^{\alpha_i}\left[\frac{\sqrt{\Lambda}\left(t_0-t\right)}{2}\right];\label{eq:cc_solution2}\\
    a_{3i}(t) &=\mathrm{exp}\left[\pm\delta_i\mathrm{exp}\left[\pm\sqrt{\Lambda}\left(t-t_0\right)\right]\right],\label{eq:cc_solution3}
\end{align}
where $a_{ni}$ are solutions of equations with $\Phi_n$. Incidentally, \eqref{eq:cc_solution2} was obtained by a different method in \cite{Meissner:1991ge}. Then, \eqref{eq:En} gives us the following conditions for the coefficients:
\begin{equation}\label{eq:negative sum of squares}
    \sum_{i}\gamma_i^2 = -1;\\
    \sum_{i}\alpha_i^2 = 1;\\
    \sum_i\delta_i^2 = 0.
\end{equation}
For solutions \eqref{eq:cc_solution1} and \eqref{eq:cc_solution3}, these conditions lead to a complex metric, and thus are not physical.

\subsection{Pre-Big Bang comments}
We briefly note a possibility of Pre-Big Bang solutions \cite{Gasperini:1991ak, Gasperini:1992em, Meissner:1991zj, Gasperini:1994xg, Gasperini:1996fu, Gasperini:2002bn}: is it possible to smoothly connect negative and positive branch around $\Phi=const.$ at $t=0$? If so, what's the evolution for $\dot\Phi \sim +0, \dot\Phi \sim -0$?\\

The answer to this question possibly lays already at the zeroth-order. Equations \eqref{first-order-dilaton-evolution} at $H=0$ , $h=0$ give vanishing $\dot\Phi$ so $\ddot \Phi = - \Lambda$. Consider a case, where the evolution starts at a point on \ref{fig:phase diagram minus zero order} and is being brought to the beginning of the coordinate system. After reaching it, we could possibly switch to the repulsive flow on \ref{fig:phase diagram plus zero order}. Value of the second derivative is determined by the sign of the cosmological constant. If the second derivative is positive (cosmological constant is negative), the gradients inverse, creating a "bouncing" universe at $H=0$ , $h=0$. \\

Similar thing may happen on the red line for the perturbative approach, avoiding the forbidden region effectively. We have verified it is possible using the corrected equivalent of \eqref{first-order-dilaton-evolution}.
\section{Solutions at higher order in $\alpha'$}
\label{sec:perturbative}
\begin{figure}[t!]
	\includegraphics[width=.475\textwidth]{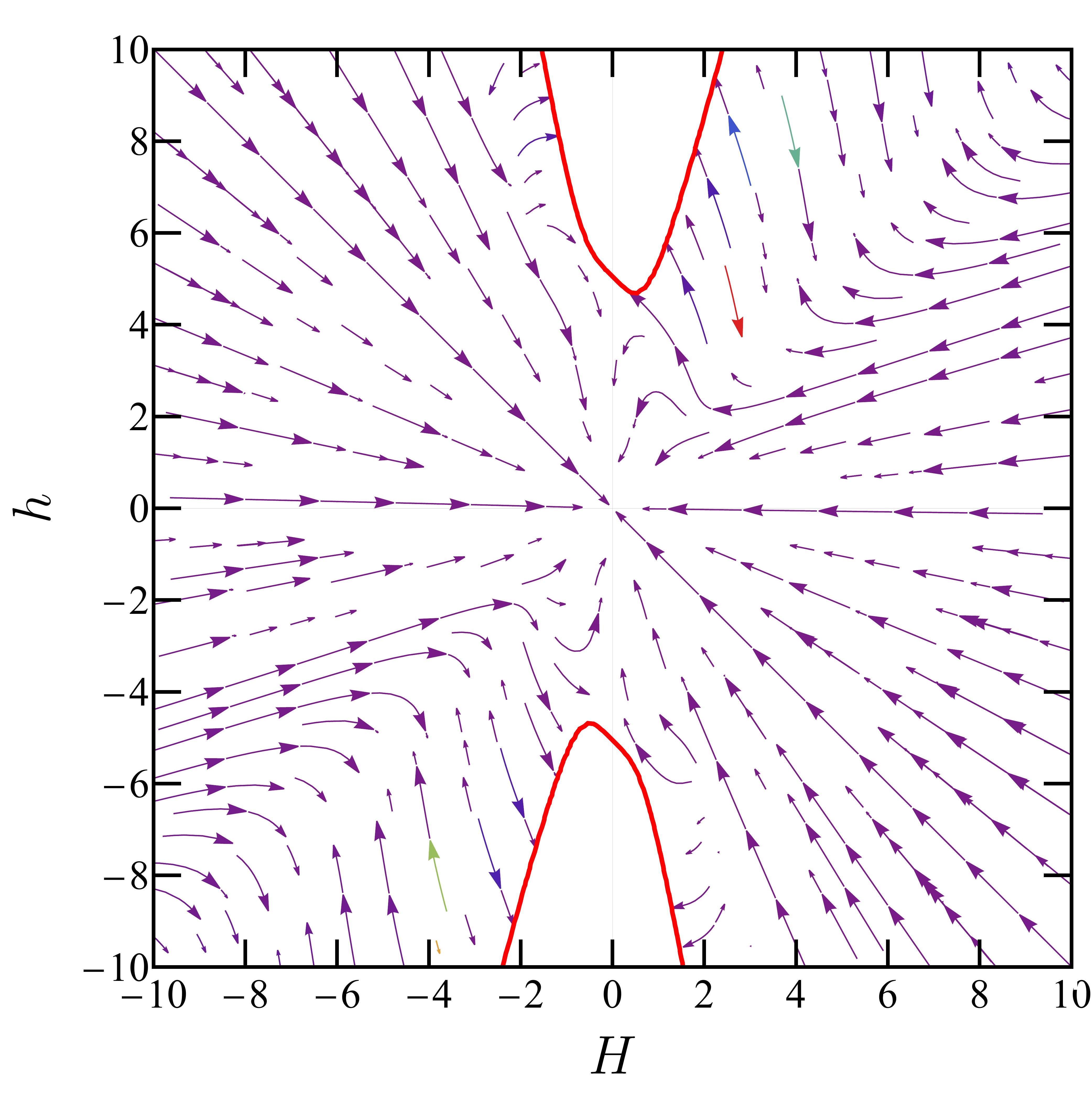}
    \caption{Phase diagram with $\dot{\Phi} < 0$ (arrows) and a curve given by $\dot{\Phi}=0$ (red) for bosonic strings.}
	\label{fig:phase diagram bosonic minus}
\end{figure}

\begin{figure}[t!]
	\includegraphics[width=.475\textwidth]{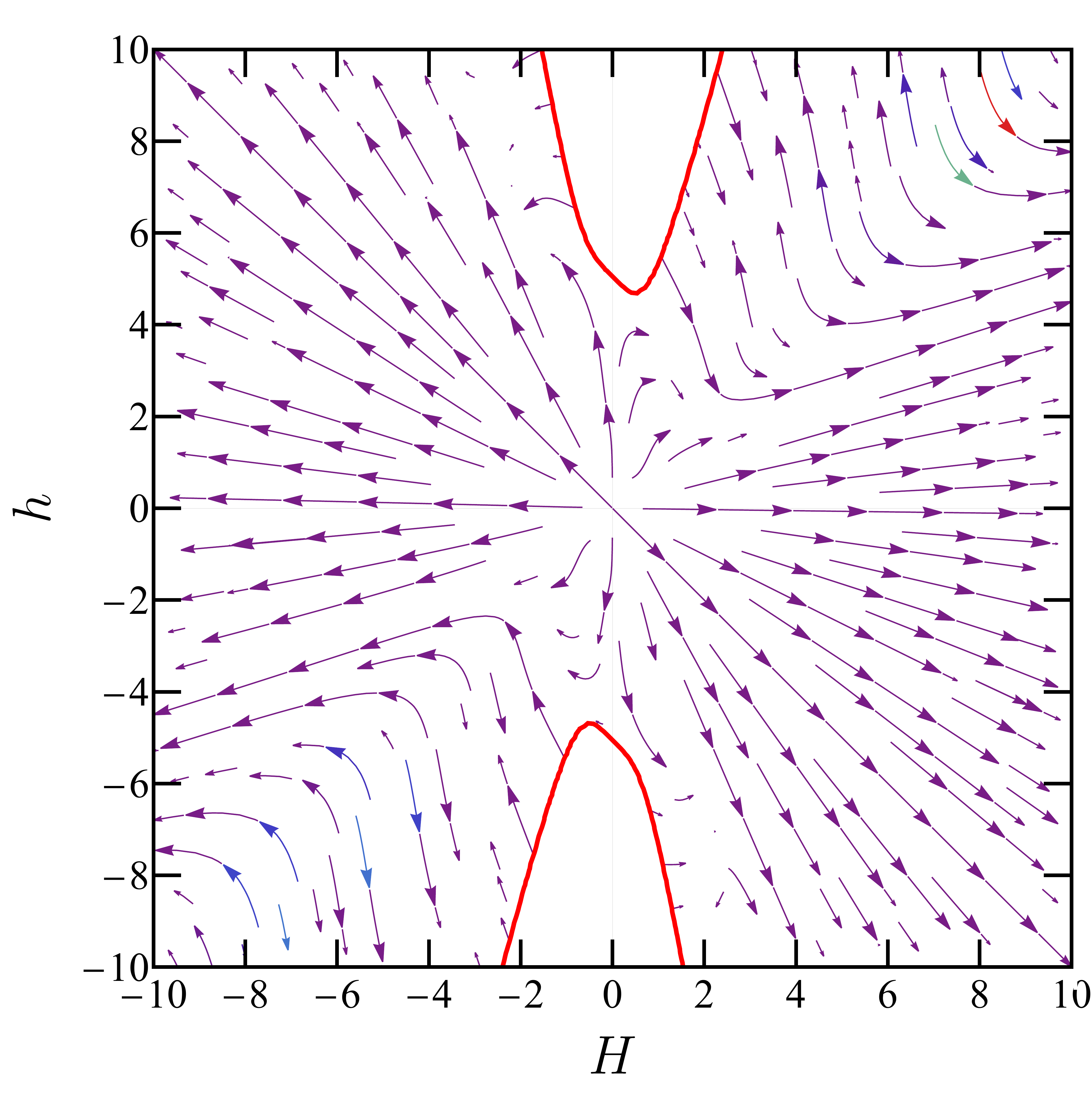}
	\caption{Phase diagram with $\dot{\Phi} > 0$ (arrows) and a curve given by $\dot{\Phi}=0$ (red) for bosonic strings.}
	\label{fig:phase diagram bosonic plus}
\end{figure}

\begin{figure}[t!]
	\includegraphics[width=.475\textwidth]{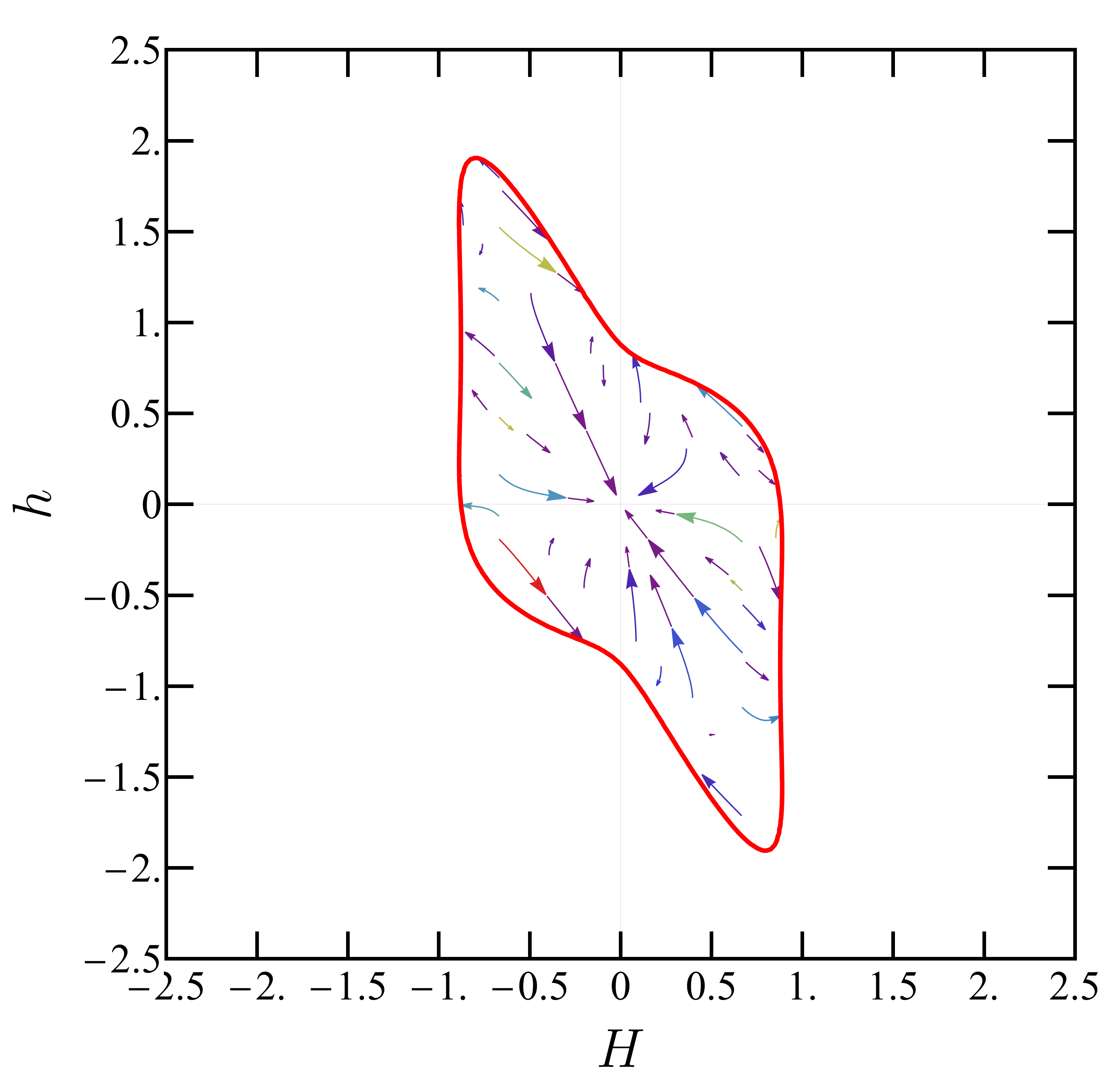}
    \caption{Phase diagram with $\dot{\Phi} < 0$ (arrows) and a curve given by $\dot{\Phi}=0$ (red) for HSZ string theory.}
	\label{fig:phase diagram hsz minus}
\end{figure}

\begin{figure}[t!]
	\includegraphics[width=.475\textwidth]{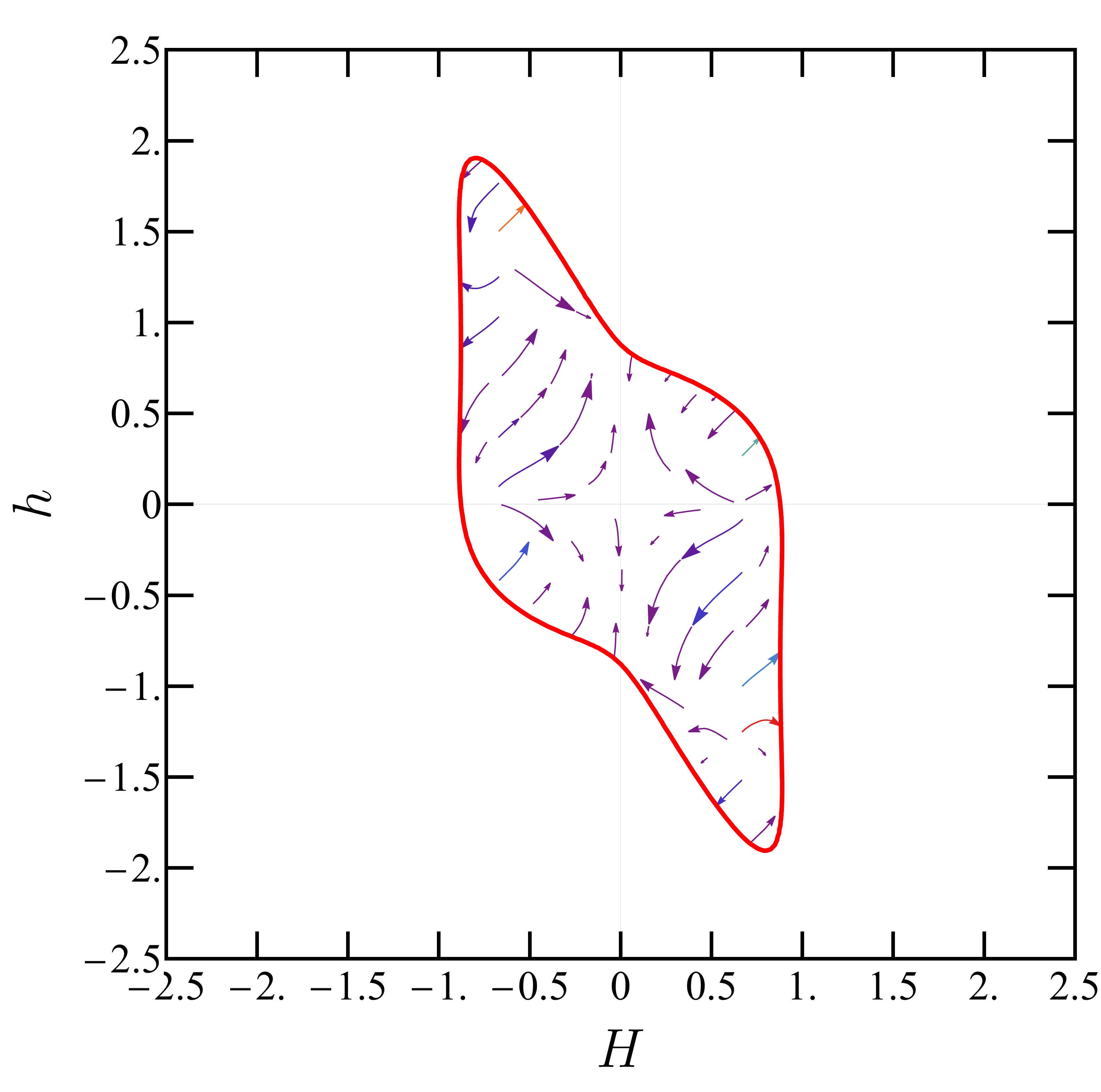}
	\caption{Phase diagram with $\dot{\Phi} > 0$ (arrows) and a curve given by $\dot{\Phi}=0$ (red) for HSZ string theory.}
	\label{fig:phase diagram hsz plus}
\end{figure}

Now, we set $H_1, \ldots,H_{d-1} = H$ and $H_d = H+h$, where $h$ is small. In this case, after using \eqref{eq:dotPhi equation} and performing the time derivative in equations \eqref{eq:Hi equations}, they become linear equations for $\dot{H_i}$ and $\dot{h}$. We solve them to obtain $\dot{H}(H,h,\mathrm{sgn}(\dot{\Phi}))$ and $\dot{h}(H,h,\mathrm{sgn}(\dot{\Phi}))$. This enables us to plot the phase diagram of this system. In table \ref{tab:c_i}, the values of the necessary constants $c_i$ in different string theories given in \cite{Codina:2021cxh} are displayed. The phase diagrams for both signs of $\dot{\Phi}$ with $d=9$, $\alpha' = 1$, $c_1=-\frac{1}{8}$ and $c_i$ as in the bosonic theory are depicted in Figures \ref{fig:phase diagram bosonic minus} and \ref{fig:phase diagram bosonic plus}, and ones with $c_i$ as in the HSZ theory are depicted in Figures \ref{fig:phase diagram hsz minus} and \ref{fig:phase diagram hsz plus}. In heterotic and type II theories, the diagrams are similar to those in bosonic theory -- conclusions drawn for the bosonic theory apply to them as well. The reason the diagram is in the HSZ theory is so different is that the coefficients at the $H^8$ terms under the square root in \eqref{eq:dotPhi equation} vanish, leaving $H^6$ as the highest power. $H^6$ terms have negative coefficients, so apart from a region near zero the expression will be negative. Therefore most of the plane is forbidden by the requirement that $\dot{\Phi}$ should be real. Further reasoning applies to theories other than HSZ.
\begin{table}[]
\centering
\label{tab:c_i}
\begin{tabular}{|c|c|c|c|c|}
\hline
          &  $c_2$&$c_3$  &$c_{4,0}$  &$c_{4,1}$  \\ \hline
Bosonic   &$\frac{1}{2^6}$  &$-\frac{1}{3\cdot2^7}$  &$\frac{1}{2^{12}}-\frac{3}{2^{12}}\zeta(3)$  &$\frac{1}{2^{16}}+\frac{1}{2^{12}}\zeta(3)$  \\ \hline
HSZ       &0  &$\frac{1}{3\cdot2^7}$  & 0 & 0 \\ \hline
Heterotic & $\frac{1}{2^7}$ &0  & $-\frac{3}{2^{12}}\zeta(3)$ & $-\frac{15}{2^{19}}+\frac{1}{2^{12}}\zeta(3)$ \\ \hline
Type II   & 0 & 0 & $-\frac{3}{2^{12}}\zeta(3)$ & $\frac{1}{2^{12}}\zeta(3)$ \\ \hline
\end{tabular}
\caption{Values of constants $c_i$ in different string theories.} 
\end{table}


Since $h$ is assumed to be a small perturbation, mainly the parts of the phase diagrams with $h \approx 0$ are of interest. One can see, that for $\dot{\Phi} < 0$, anisotropic perturbations are suppressed regardless of the value of $H$, but $H$ tends to 0 (which corresponds to a static universe). For $\dot{\Phi} > 0$, anisotropic perturbations will grow for any value of $H$. For other theories, the diagrams are similar -- the conclusions stay the same. 
The white regions on the top and bottom of both diagrams correspond to the values of $H$ and $h$ for which $\dot{\Phi}^2<0$ and thus are prohibited. When $\dot{\Phi}=0$, $\dot{H}=\dot{h}=0$ and \eqref{eq:dotPhi equation} becomes an algebraic constraint. It is satisfied for $H=h=0$ and on the curves marked in red on the diagrams. Change of sign of $\dot{\Phi}$ is possible only there (as we expect $\dot{\Phi}$ to be continuous). When $\dot{\Phi}>0$, the arrows on the phase diagram point away from 0 and the curve, and when $\dot{\Phi}<0$, they point towards 0 and the curve. Therefore, this change of sign is likely only possible in one direction -- from negative to positive.

\begin{figure}[t!]
	\includegraphics[width=.475\textwidth]{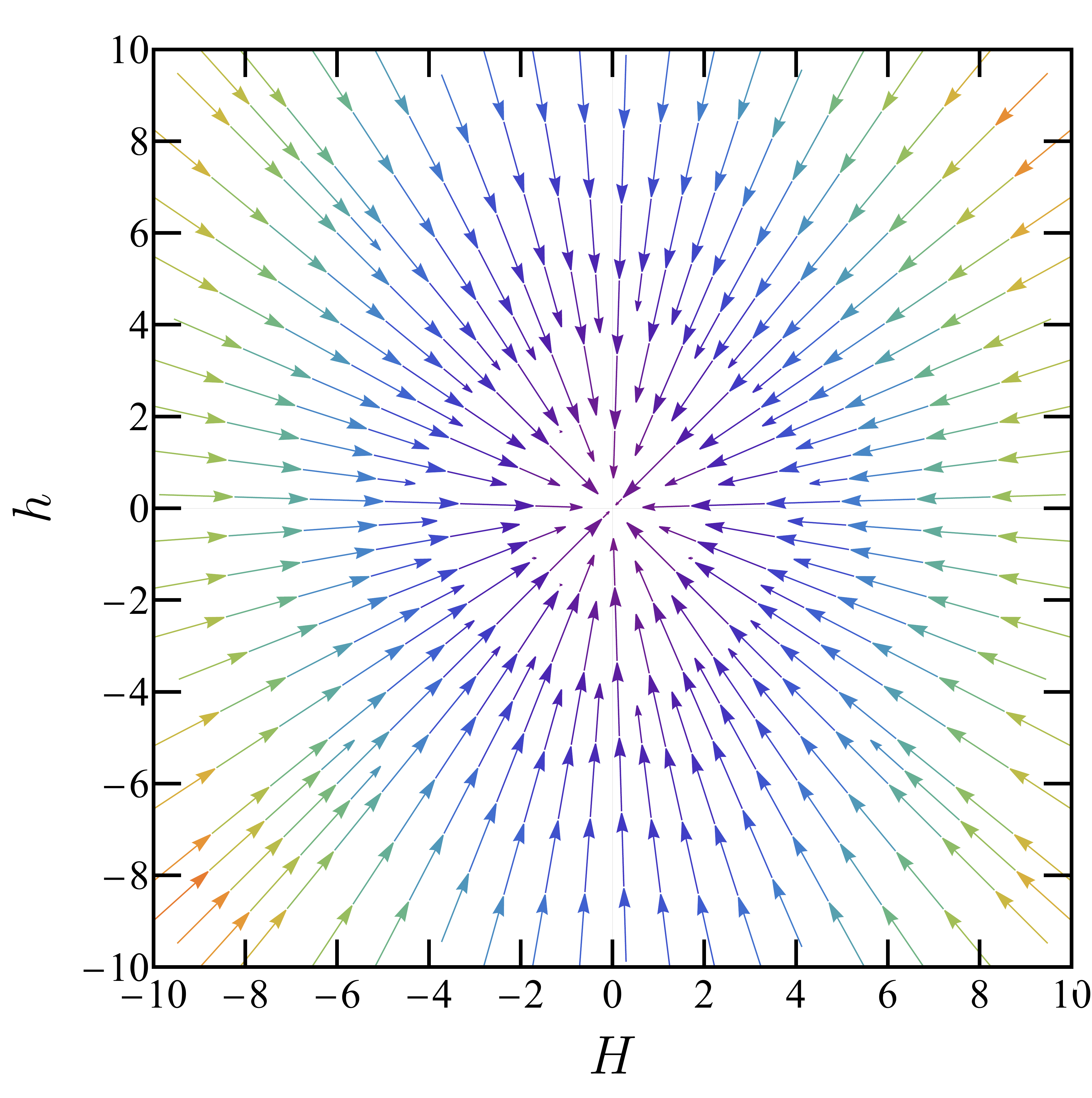}
    \caption{Phase diagram with $\dot{\Phi} < 0$ at zero order in $\alpha'$.}
	\label{fig:phase diagram minus zero order}
\end{figure}

\begin{figure}[t!]
	\includegraphics[width=.475\textwidth]{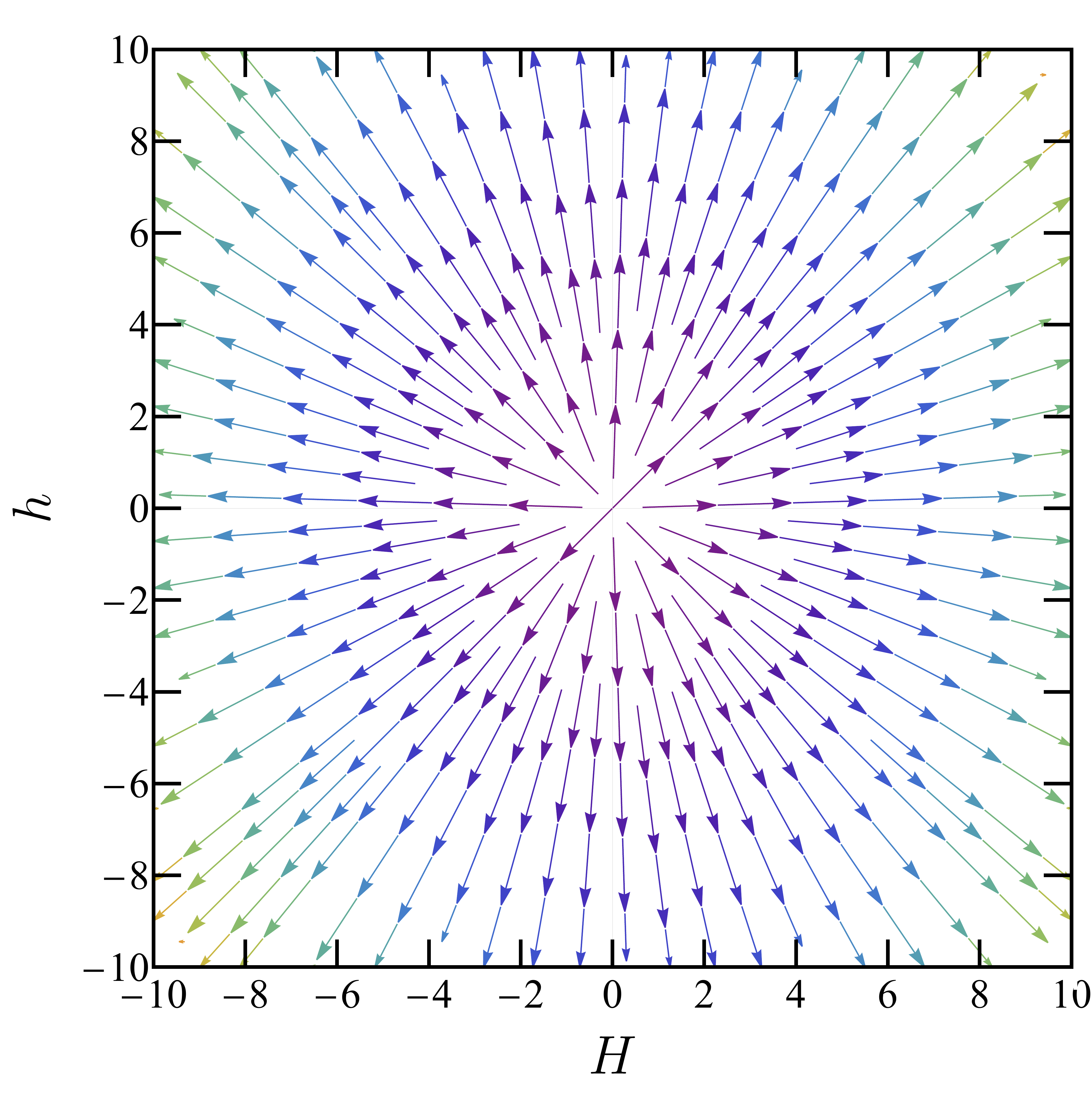}
    \caption{Phase diagram with $\dot{\Phi} > 0$ at zero order in $\alpha'$.}
	\label{fig:phase diagram plus zero order}
\end{figure}

At the zeroth order in $\alpha'$, analogously derived equations are:
\begin{equation}\label{eq:dotH anisotropic perturbation}
	\dot{H} = \left(\mathrm{sgn}(\dot{\Phi}) \sqrt{dH^2+2hH+h^2}\right)H,
\end{equation}
\begin{equation}\label{eq:dotH+h anisotropic perturbation}
	\dot{H} + \dot{h} = \left(\mathrm{sgn}(\dot{\Phi}) \sqrt{dH^2+2hH+h^2}\right)(H+h);
\end{equation}
which lead to phase diagrams depicted in Figures \ref{fig:phase diagram minus zero order} and \ref{fig:phase diagram plus zero order}. Near the $H$ axis, the diagrams at $O(1)$ and $O(\alpha'^3)$ are mostly identical, which means that for initial conditions in this regions the solutions will be similar to those described in the previous section. The most significant difference is the appearance of the forbidden regions and the possibility of change of sign of $\dot{\Phi}$. As one can see in \eqref{eq:Phi interesting solution}, at the zeroth order the sign of $\dot{\Phi}$ is determined by the choice of initial conditions. 

\section{Conclusions}
Our investigation reveals that the de-Sitter solution (an isotropic one with constant $H$ greater than 0), that to the third order in $\alpha'$, no stable solution of this kind is possible, because either the only stable point is $H=0$ (for $\dot{\Phi}<0$), or there are no stable points. This confirms the previous studies done at $\alpha'{}^0$.\\ 
However, no conclusions can be drawn about the stability of non-perturbative de-Sitter vacua, since higher-order terms can change the dynamics of the system. However, our analysis entails that 
for small values of $H$ the phase diagrams at $\alpha'{}^0$ and $\alpha'{}^3$ do not differ significantly supporting the claim  for the lack of de-Sitter solutions at the observed value of cosmological constant \cite{Obied:2018sgi, Garg:2018reu, Ooguri:2018wrx}.
Yet, for the non-perturbative, isotropic study see our previous work \cite{Bieniek:2022mrv}.\\
Furthermore, we have discussed a possible Pre-Big Bang solution in a universe with negative cosmological constant.
\addcontentsline{toc}{section}{The Bibliography}
\acknowledgments
J.C. thanks Robert Brandenberger and Heliudson Bernardo for numerous engaging discussions on the stability of de Sitter spacetime. J.C. is thankful for the hospitality of the McGill University.
J.H.K thanks Massachusetts Institute of Technology (MIT) for hospitality and support during this work. J.H.K. was supported by the Polish National Science Centre grant 2018/29/N/ST2/01743. K.A.M. and J.H.K. were partially supported by the Polish National Science Centre grant UMO-2020/39/B/ST2/01279.
\addcontentsline{toc}{section}{The Bibliography}
\bibliography{References.bib}{}
\bibliographystyle{apsrev4-1}
\end{document}